\documentclass[journal abbreviation, manuscript]{copernicus}
\AtBeginDocument{\fontsize{11}{18}\selectfont}
\usepackage{setspace}
\usepackage{xspace}
\usepackage{hyperref}
\usepackage[noabbrev]{cleveref}
\usepackage{booktabs}     
\usepackage{tabularx}    
\usepackage{xltabular}    
\usepackage{multirow}     
\usepackage{ragged2e}   
\usepackage{rotating}   
\usepackage{pdflscape}   
\usepackage[table]{xcolor}
\usepackage{url}

\usepackage{setspace}
\definecolor{rPos}{RGB}{27,120,55}
\definecolor{rNeu}{RGB}{120,120,120}
\definecolor{rNeg}{RGB}{179,45,45}
\newcommand{\rfull}{$\blacksquare$}  % Essential  = \textcolor{rPos}{\textbf{3}}
\newcommand{\rmost}{$\boxtimes$}  % Important = \textcolor{rPos}{\textbf{2}}
\newcommand{\rpart}{$\square$}  % Nice to have = \textcolor{rNeu}{\square}
\newcommand{\rwk}{$\cdot$}  % Irrelevant = 0
\usepackage{glossaries}

\begin{document}
\nolinenumbers

\title{\textbf{A Framework for AI-Ready Data in Earth and Environmental Science}}

% \Author[affil]{given_name}{surname}

\Author[1][tobias.weigel@hereon.de]{Tobias}{Weigel} %% correspondence author
\Author[2,3,4]{Jan}{Bumberger}
\Author[1]{Danu}{Caus}
\Author[5]{Christian}{Chwala}
\Author[5]{Romy}{Fösig}
\Author[6]{Sonja}{Hänzelmann}
\Author[5]{Sibylle K.}{Hassler}
\Author[7]{Beate}{Hetenyi}
\Author[8]{Ralf}{Kunkel}
\Author[5]{Katharina}{Loewe}
\Author[8]{Ulrich}{Loup}
\Author[8]{Yueling}{Ma}
\Author[6]{Stefan}{Neuhaus}
\Author[5]{Kaleb}{Phipps}
\Author[8]{Tino}{Polen}
\Author[6]{Maren}{Rebke}
\Author[1]{Björn L.}{Saß}
\Author[2,3]{David}{Schäfer}
\Author[9]{Timm}{Schoening}
\Author[9]{Emanuel}{Söding}
\Author[7]{Paula}{Stock}

\affil[1]{Helmholtz-Zentrum Hereon, Max-Planck-Straße 1, 21502 Geesthacht, Germany}
\affil[2]{Helmholtz Centre for Environmental Research -- UFZ, Research Data Management -- RDM, Permoserstraße 15, Leipzig, 04318, Germany}
\affil[3]{Helmholtz Centre for Environmental Research -- UFZ, Department Monitoring and Exploration Technologies, Permoserstraße 15, Leipzig, 04318, Germany}
\affil[4]{German Centre for Integrative Biodiversity Research (iDiv) Halle-Jena-Leipzig, Puschstraße 4, Leipzig, 04103, Germany}
\affil[5]{Karlsruhe Institute of Technology, Kaiserstraße 12, 76131 Karlsruhe}
\affil[6]{Alfred Wegener Institute, Helmholtz Centre for Polar and Marine Research}
\affil[7]{GFZ Helmholtz Centre for Geosciences, Telegrafenberg, 14473 Potsdam}
\affil[8]{Forschungszentrum Jülich}
\affil[9]{GEOMAR Helmholtz Centre for Ocean Research}

%% The [] brackets identify the author with the corresponding affiliation. 1, 2, 3, etc. should be inserted.

%% If an author is deceased, please add \deceased[$Deceased date if applicable$]{$Author number$} (e.g. \deceased[13 November 2015]{2}) at the end of the affiliations. The author number depends on the placement of the author in the author list, e.g. the third author has number 3.

%% If authors contributed equally, please add \equalcontrib{$Author numbers$} (e.g. \equalcontrib{1,3}) at the end of the affiliations. The author number depends on the placement of the author in the author list, e.g. the third author has number 3.

\runningtitle{TEXT}

\runningauthor{TEXT}

\received{}
\pubdiscuss{} %% only important for two-stage journals
\revised{}
\accepted{}
\published{}

%% These dates will be inserted by Copernicus Publications during the typesetting process.

\firstpage{1}

\maketitle

\begin{abstract}

Artificial intelligence (AI), specifically machine learning (ML), is increasingly integral for Earth and environmental research, yet the suitability of current research data infrastructures for supporting AI workflows remains poorly defined. While the FAIR principles provide a widely adopted foundation for scientific data management, they do not explicitly address several requirements emerging when datasets are used for training, evaluating, and deploying AI models. This article presents a practically motivated framework for assessing and improving AI-readiness of research data within distributed Earth and environmental data infrastructures. The framework was developed through a structured, community-driven process, initiated with a review of existing literature and readiness frameworks combined with feedback from a limited survey of scientists applying machine learning methods. Its main aspects were shaped by a multidisciplinary expert group representing data providers, infrastructure developers, data managers, and AI practitioners. The framework identifies 25  criteria covering data preparation, data quality, documentation, access, and supervised learning targets. The proposed criteria emphasize aspects that are particularly relevant for AI applications, including provenance, treatment of missing data, bias documentation, uncertainty characterization, machine-actionable metadata, efficient access formats, and the quality and reproducibility of target generation. The framework is applied to a broad range of representative use cases from environmental sensor quality control and global weather forecasting to marine biology, highlighting how the relevance of individual criteria varies depending on the application. We argue that AI-readiness should be understood primarily as a property of transparency and fitness for purpose rather than as a fixed level of data quality. The framework lays a foundation for communication and community-driven action to advance AI-ready data and support next-generation AI-enabled scientific workflows.

\end{abstract}

%\copyrightstatement{TEXT} %% This section is optional and can be used for copyright transfers.

\introduction  %% \introduction[modified heading if necessary]

The FAIR data principles \citep{Wilkinson2016,rda2020fair} form an established conceptual framework for managing scientific data, underpinning procedures that aim to ensure that their quality meets stakeholder requirements and that their value is optimized. Artificial Intelligence (AI) methods, specifically Machine Learning (ML), are fostering large-scale innovations across scientific disciplines. It is widely recognized that employing ML, Deep Learning in particular, requires large amounts of data and benefits from assured data quality \citep{hestness2017}. However, there is still a disconnect between processes and stakeholders responsible for data management and those focusing on ML development. While the FAIR principles offer a solid framework for scientific data management, they fail to meet significant ML user needs in practice. In this article, we introduce a conceptual framework for AI-readiness of data that was developed in the context of cross-institutional workflows in Earth and Environment research. The goal of the framework is to clarify and acknowledge user requirements, stimulate community dialogue, facilitate the development of guidelines and enable communication across workflow stages.

Data infrastructures are tasked with enabling downstream AI workflows. However, significant ambiguity remains concerning what steps should follow harmonizing metadata and implementing FAIR principles, when moving towards AI workflows. We often observe that current procedures break down at the point of depositing data into a repository or delivering data streams. Data infrastructure architects often lack insight into how data is actually used \citep{CutcherGershenfeld201688}. The viewpoint of AI users, in particular, is unclear and rarely understood \citep{sambasivan2021}.

The aim of this article is to activate the relevant communities and bridge the gap in understanding between data infrastructure and data use with respect to AI applications, thereby establishing a foundation for communication. We achieve this by providing a framework of essential criteria for AI-readiness along exemplary use cases. These criteria have the potential to inform the development of community-led, generally applicable guidelines for data infrastructures that can improve AI-readiness. It must also be emphasized that no effort at the data infrastructure level can and should fully eliminate the need for ML users to explore data.

Typical data infrastructures involve the following three key stakeholders. \emph{Data providers} are the primary source of knowledge of how data were recorded or generated. \emph{Data managers} are responsible for codifying and curating this knowledge and making it available for further use. Finally, \emph{ML developers} use data and rely on recorded knowledge to understand and correctly represent the context in which the data were created.

ML developers are a particularly important stakeholder here as they represent the user perspective. ML methods, specifically the class of algorithms associated with Deep Learning, are currently the main driver of widely perceived innovations across science and society. ML algorithms are designed to learn a function or relationship directly from large amounts of training data. Supervised ML algorithms also require \emph{targets}, e.g., not just the image of a cat, but also a text label that states that the image shows a cat. In contrast, a typical unsupervised algorithm would be to perform the task without a specific label, e.g. clustering similar images based solely on the data contained within the data. While there are many different algorithms, their practical application often requires an understanding of the properties of the data, such as biases, over-/under-sampling, noise, errors and gaps. Understanding these properties requires investigation into upstream data processing workflows and general data provenance, potentially including details of the measurement procedures or instruments in case of observational data. This is the scope of data exploration, which is an early essential step in developing ML applications, and a later step when fine-tuning and debugging models to improve their performance. 

Typically, FAIR data workflows lack some of the details that are essential for applying ML methods. Moreover, even small structural differences in the way data are encoded or formatted can be a major obstacle to applying ML algorithms across datasets. Even in commonly perceived well-curated projects, minor differences can turn out to be a hassle for ML workflows \citep{VanKatwyk2026}.

AI-readiness transcends the more general readiness of data for analysis by emphasizing several elements that are characteristic for ML workflows. These include the dedicated capture of knowledge about targets for supervised learning, an emphasis on describing biases, gaps and data quality criteria such as measurement errors or uncertainties, and an emphasis on making such information accessible to libraries and tools in standardized form in order to foster automatization and lowering the cost of developing workflows that require heterogeneous input datasets.

There are some key requirements a framework for AI-readiness must follow. It should be aligned and compatible with the FAIR data principles as setting up a conflicting framework would be highly counterproductive. At the same time, recommendations must strike a good balance between exhaustiveness and practicality. Curating datasets for all possible downstream uses is a futile effort as the applications are highly variable and \textit{a priori} unknown. They can only be estimated, based on commonly accepted typical cases with potentially contradicting priorities. For example, for one given ML application, gaps in the data can be undesirable as they make the application of certain algorithms impossible or difficult (e.g., requiring additional interpolation), while in other cases, intelligent filling of gaps or detection and correction of faulty samples is the actual ML application to be developed.

Therefore, the framework should focus on common aspects across use cases. It cannot address all downstream questions, as doing so would potentially exceed the capacities of the upstream data documentation workflows. Instead, we focus on providing a set of essential elements in the form of a list of criteria for AI-readiness.

In the following, \cref{sec:methods} describes the motivational background for our work. \Cref{sec:related} details the results of an initial landscaping exercise, surveying literature and ongoing or emerging efforts to review FAIR data principles in light of ML usage. 
In \cref{sec:criteria}, the AI-readiness criteria are presented and discussed. Their development was informed by the state of the art, survey results and feedback from data providers and data infrastructure builders. 
\Cref{sec:use-case} describes a set of example use cases that illustrate the interpretation of criteria in practice.
The discussion in \Cref{sec:discussion} highlights trade-offs and open points.

\section{Background}
\label{sec:methods}

Since summer 2025, a representative group of experts has investigated AI-readiness of data in the scope of an environmental data infrastructure. The group members represent manifold stakeholder interests, including subdomains of the research field and both modeling and observation. Data provider and infrastructure concerns as well as research groups that use ML methods on data are represented. Work on the framework was conducted through at least semi-monthly meetings, with consensus established through explaining motivations and balancing interests of the stakeholder groups, working towards a common understanding. 

The immediate motivational case for an infrastructure in the Earth sciences is the Helmholtz Earth and Environment DataHub, a distributed, multi-institutional data space \citep{nagel_lycklama_2021} managed by a consortium of research centers in the Helmholtz Research Field Earth and Environment. As a typical infrastructure in the Earth sciences, its data cover many Earth science sub-disciplines and modalities, including observation data products, time series, environmental samples, and digitally generated data from model simulation runs. The data workflows are designed to gather information across institutions. As a collaborative, cross-institutional infrastructure, the DataHub benefits from strategic institutional backing to develop services and to implement consistent procedures across institutions, which is essential for defining and implementing AI-readiness. The DataHub is one example within a wider landscape of distributed Earth and environmental research infrastructures, including NFDI4Earth at the national level \citep{Bernard2023}, European research infrastructures that also contribute to the ENVRI community \citep{Petzold2019,Petzold2024}, and the global GERI network \citep{Loescher2022}. Although these initiatives differ in scope and technical implementation, they share the challenge of translating FAIR commitments into cross-institutional data practices and services that support downstream analysis and AI.

In order to initiate work on relevant criteria, we conducted a limited survey among immediate colleagues who develop ML applications. While the survey was not extensive enough to fully review criteria (only 10 responses were eligible for evaluation), it highlighted concrete needs and informed the weighting of criteria in the subsequent discussions. For instance, the outcomes identified targets as critical objects for quality control and curation, while underscoring the importance of providing datasets that can be accessed efficiently using widely adopted software libraries. In addition, they demonstrated that substantial optimization of machine learning model performance often depends on detailed knowledge of data provenance and a thorough understanding of the procedures through which the data were generated or observed.  Further details on the survey are given in appendix \ref{sec:appendix-survey}.

\section{Related Work}
\label{sec:related}

Transparent and well-documented data are prerequisites for robust and reproducible research. The FAIR principles provide a widely adopted foundation by requiring data to be findable, accessible, interoperable, and reusable \citep{Wilkinson2016}. Initiatives such as the FAIR Data Maturity Model \citep{rda2020fair} and the Cross-Domain Interoperability Framework\footnote{\url{https://cross-domain-interoperability-framework.github.io/cdifbook/}, accessed 2026-09-17} translate these principles into assessment criteria and interoperability practices. The Croissant framework \citep{Akhtar2024} provides a metadata vocabulary for recording information across ML datasets and enabling ML workflows to use them. These approaches are deliberately application-agnostic, however, and FAIR compliance alone does not establish that a dataset can be used efficiently and reliably in a machine-learning workflow. Such workflows may additionally require machine-actionable provenance, consistent encoding, explicit descriptions of missingness, bias, and uncertainty, efficient access patterns, and, for supervised learning, documented targets. In consequence, the FAIR for AI initiative frames AI-readiness as a complementary concern rather than a replacement for FAIR \citep{huerta2023fair}.

Several frameworks provide a shared vocabulary for assessing this additional readiness. Data Readiness Levels distinguish accessibility, faithfulness, and contextual appropriateness \citep{lawrence2017datareadinesslevels}, while the ABC framework organizes readiness into progressively more demanding tiers ranging from raw data collection to task-specific, ethical, and legal suitability \citep{castelijns2020}. More recent taxonomies and actionable frameworks cover data quality, understandability, organization, bias, governance, and the effect of data properties on AI systems \citep{Hiniduma2025,Majithia2026}. Tools such as the AI Data Readiness Inspector further seek to quantify these dimensions and partly automate their assessment \citep{Hiniduma2024}. These approaches make readiness more explicit and comparable, but a single level or aggregate score can obscure task dependence. A property that is beneficial for one application may be unnecessary or even undesirable for another; for example, pre-filling data gaps may facilitate some models while keeping and documenting them is required to develop and evaluate an imputation method.

Other initiatives approach AI-readiness by extending FAIR practices or implementing them within particular domains. AI-READI applies FAIR-oriented requirements to biomedical datasets\footnote{\url{https://docs.aireadi.org/docs/3/AI-readiness/fair-principles}, accessed 2026-09-17}, whereas FAIR2\footnote{\url{https://www.fair2.ai}, last accessed 2026-09-17} emphasizes context-rich, machine-actionable metadata and alignment with responsible AI. Work on scientific AI at scale additionally highlights the operational importance of data layout, access, and movement when datasets must support large distributed workflows \citep{Brewer2025}. These efforts demonstrate that AI-readiness depends not only on intrinsic data quality, but also on documentation, governance, representation, and the infrastructure through which data are accessed.

For environmental data, the NOAA readiness levels\footnote{\url{https://www.star.nesdis.noaa.gov/star/documents/meetings/2020AI/presentations/202010/20201022_Christensen.pdf}, last accessed 2026-09-17} and the ESIP Data Readiness Cluster checklist are closest to the presented work \citep{esip2023checklist}. The NOAA readiness levels distinguish between data quality, access and documentation, but do not define more detailed criteria and level definitions. The ESIP checklist covers data preparation, quality, documentation, and access, and has been operationalized through an interactive documentation tool by the UK Met Office\footnote{\url{https://github.com/MetOffice/ai_data_readiness}, last accessed 2026-09-17}. Its breadth makes it a valuable starting point, but some questions are formulated as prescribed transformations rather than transparency requirements. Asking whether all gaps have been filled, for example, does not distinguish between applications that require complete inputs and those for which gap detection or imputation is the learning task.  

Existing work thus provides general principles, maturity levels, domain-specific implementations, quantitative assessments, and operational checklists. Yet, what is missing is a framework that covers the concerns of all stakeholder groups and balances level of detail with generalizability. Given our motivation, it should also take into account requirements specific to the Earth sciences where necessary. The framework developed here builds on these foundations while treating AI-readiness primarily as a matter of transparency and fitness for purpose. 

\section{AI-Readiness Criteria}
\label{sec:criteria}
\label{sec:criteria-discussion}

\begin{table}[t]
  \centering
  \footnotesize
  \setlength{\tabcolsep}{4pt}
  \renewcommand{\arraystretch}{1.3}
  \caption{Short descriptions of the AI-readiness criteria. *Asterisks mark criteria that are identical to FAIR criteria \citep{Wilkinson2016}. The last column indicates rough provenance with the most suitable ESIP
    criterion~\citep{esip2023checklist}; see reference table in \Cref{sec:appendix-esip}. Each
    criterion carries an AIR-ID code for referencing throughout this manuscript. They are grouped into five categories: data preparation (P), data quality (Q), data documentation (D), data access (A), targets (T).}
  \label{tab:air-overview}
  \begin{tabularx}{\linewidth}{@{}c l >{\RaggedRight\arraybackslash}X c@{}}
    \toprule
    & \textbf{AIR-ID} & \textbf{Criterion (short description)} & \textbf{ESIP ID} \\
    \midrule
 
    \multirow{6}{*}{\rotatebox[origin=c]{90}{{\scriptsize\bfseries Data preparation (P)}}}
      & P-1.1            & Missing values and data gaps have been quantified, documented, and either imputed with declared methods or explicitly left as missing. & 1.1 \\
      & P-1.2            & Outliers and corrupt values are detected, quantified, and documented. & 1.2 \\
      & P-1.3            & Duplicate data are flagged or removed and the process is documented. & --- \\
      & P-2.1\textsuperscript{*} & The provenance of processing steps, tools and parameters is documented. & 2.8 \\
      & P-2.2            & The level of data processing is explicitly defined and documented, including distinctions between raw and derived data. & 2.11 \\
      & P-3              & The human operator decisions during data preparation are documented. & --- \\
    \midrule
 
    \multirow{4}{*}{\rotatebox[origin=c]{90}{{\scriptsize\bfseries Quality (Q)}}}
      & Q-1.1            & The data include all relevant available data points, and any excluded data are documented along with the reasons for their exclusion. & 2.1 \\
      & Q-1.2            & The sources of bias are identified, assessed, and documented, along with potential mitigation strategies. & 2.3 \\
      & Q-2              & The metadata contains quantitative measures of uncertainty. & 2.5 \\
      & Q-3              & The metadata contains quantitative information of data resolution and extent in space and time. & 2.6 \\
    \midrule
 
    \multirow{7}{*}{\rotatebox[origin=c]{90}{{\scriptsize\bfseries Documentation (D)}}}
      & D-1.1\textsuperscript{*} & The data are accompanied by rich metadata. & 3.1 \\
      & D-1.2\textsuperscript{*} & The metadata is machine-readable. & 3.1.2 \\
      & D-1.3\textsuperscript{*} & The metadata is expressed in compliance with a machine-actionable community standard. & 3.1.1 \\
      & D-2.1\textsuperscript{*} & The data dictionary is machine-readable and encoded using an open format. & 3.2.2 \\
      & D-2.2\textsuperscript{*} & The dataset variables and parameters are described by a comprehensive data dictionary. & 3.2 \\
      & D-2.3\textsuperscript{*} & The data dictionary entries follow standardized vocabularies and schemas. & 3.2.1 \\
      & D-3              & The loading scripts are available and documented. & --- \\
    \midrule
 
    \multirow{5}{*}{\rotatebox[origin=c]{90}{{\scriptsize\bfseries Access (A)}}}
      & A-1.1\textsuperscript{*} & The data are accompanied by a clear data usage license. & 3.7 \\
      & A-1.2\textsuperscript{*} & The data usage license is standardized and machine-readable. & 3.7.1 \\
      & A-2              & The metadata provides information on the consent of data creators to allow AI-based exploitation. & --- \\
      & A-3.1\textsuperscript{*} & The data are available in at least one open, non-proprietary file format. & 4.1.2 \\
      & A-3.2\textsuperscript{*} & The data file format affords ease of use and efficiency in ML workflows. & 4.1.1 \\
    \midrule
 
    \multirow{3}{*}{\rotatebox[origin=c]{90}{{\scriptsize\bfseries Targets (T)}}}
      & T-1.1            & If targets are available, they are provided in an open format. & 1.5 \\
      & T-1.2            & The targets were subject to a documented QA/QC process and the process for generating targets is documented. & --- \\
      & T-1.3            & The labels follow a controlled vocabulary or ontology. & --- \\
    \bottomrule
  \end{tabularx}
\end{table}

Based on related prior work, survey results, and expert opinions from the stakeholders in the working group, we developed a list of criteria summarized in Table~\ref{tab:air-overview}. For each criterion, the table indicates relation to the FAIR data principles and the ESIP checklist. A more detailed explanation of each criterion is given in the appendix in Table~\ref{tab:air-detail}. Note that in many cases, while there is a rough provenance to the ESIP criteria, the criteria presented here have undergone significant changes, reducing ambiguity in their formulation and sharpening them in light of concrete stakeholder needs. 

While the criteria are stated as binary requirements, in practice, the extent to which they are or can be fulfilled will be the main scale, with complete fulfillment being likely a rare case. This scale of fulfillment needs to be distinguished from the scale of importance of the criterion, which can only be assessed against a concrete use case. Example applications are given in section \ref{sec:use-case}. This also raises the question of who assesses fulfillment or importance. The framework can be used for case specific self-assessment, while future work could establish independent certification.

The criteria presented in Table~\ref{tab:air-overview} are intended as a practical framework for improving the transparency of datasets with respect to their potential use in AI and in particular machine learning workflows. They do not prescribe specific implementation approaches, metadata schemas, or infrastructure solutions. Rather, they define the information that users of data require in order to assess whether a dataset is suitable for AI-based applications. In this sense, the criteria represent user-oriented transparency requirements that can guide the development of metadata standards, data management practices, and infrastructure services.

A central design principle of the criteria is that AI-readiness is not equivalent to data quality in a narrow sense. Instead of introducing a single criterion stating that a dataset is of “high quality”, the framework decomposes quality-related concerns into a set of observable and documentable characteristics. Examples include the handling of missing values (P-1.1), outlier detection (P-1.2), documentation of duplicate records (P-1.3), completeness of available observations (Q-1.1), and the disclosure of uncertainties and biases (Q-1.2, Q-2). This approach reflects the observation that different AI use cases may place different importance on particular quality dimensions. By requiring transparency about these aspects rather than imposing universal thresholds, the framework enables users to make informed decisions about fitness for purpose.

The criteria also intentionally overlap with established FAIR principles. Many of the documentation and access criteria correspond to FAIR requirements related to metadata richness, machine readability, standardization, licensing, and accessibility. However, FAIR compliance alone does not necessarily provide the information required to evaluate suitability for AI workflows. The criteria therefore complement existing FAIR-oriented approaches by extending them with AI-specific concerns such as bias documentation, uncertainty quantification, treatment of missing data, provenance of processing steps, and documentation of target generation. At the same time, the framework avoids assuming prior FAIR compliance, allowing it to be applied as a standalone checklist while remaining compatible with FAIR implementation efforts.

Several criteria are organized hierarchically to reflect increasing levels of machine usability. This is particularly visible in the metadata and documentation category. Criterion D‑1.1 requires the existence of rich metadata, D‑1.2 requires that metadata is machine-readable, and D‑1.3 further requires compliance with machine-actionable community standards. A similar progression is evident for data dictionaries and licensing information. This layered structure acknowledges that information may exist without being readily accessible to automated systems, while AI workflows increasingly depend on metadata that can be discovered, interpreted, and acted upon by software agents with minimal human intervention.

The inclusion of a dedicated category on targets reflects the importance of supervised machine learning workflows, where target categories or variables strongly influence model performance and interpretability. The criteria in this category address the availability, quality assurance, and standardization of targets when such information exists. However, targets occupy a different position within the data life cycle than many of the other criteria as targets are often created post-hoc. In numerous real-world scenarios, datasets were not originally created for AI applications, and the generation of categorical labels or target variables often occurs later as part of a separate data science process (e.g., when manually creating labels for existing imaging datasets). Data providers are not responsible for creating targets, as this task is done by data scientists, annotators and domain experts. The target-related criteria should be understood as transparency requirements when targets are available rather than a requirement that data providers create them.

This distinction highlights a broader shift in responsibility when existing datasets are repurposed for AI. AI-readiness becomes the joint responsibility of those who generate data or targets and those who quality-control, document, distribute and curate them. In many cases, past processes at early stages of the data life cycle, such as data acquisition and initial processing, can no longer be altered. AI-readiness therefore becomes less a question of producing perfect data and more a question of making limitations, assumptions, and potential risks visible. Responsibility shifts from the original act of data creation toward the interpretation, documentation, governance, and accountable reuse of data. From this perspective, AI-readiness is fundamentally linked to transparency rather than perfection.

One topic that emerged repeatedly during the development of the framework is the provision of ethical and legal information. Ethical considerations, data anonymization procedures, consent restrictions, genetic resource agreements, and limitations related to sensitive or protected locations can all influence whether data are suitable for AI applications. Furthermore, such measures may themselves introduce biases that affect downstream models. Although the working group considered including a dedicated criterion covering legal and ethical information, the diversity and context dependence of these issues made it difficult to formulate a single criterion applicable across scientific domains. Instead, the current framework includes creator consent for AI-based exploitation (A‑2) while identifying the broader ethical and legal dimension as an important area for future refinement.

\section{Exemplary Use Cases}
\label{sec:use-case}

We assess the context dependence of the proposed criteria using applications from Earth and Environment research. Table~\ref{tab:air-usecases} reports the importance of each criterion to the AI component of the individual use case, assessed from a user perspective (the scientists who performed the corresponding case study). Detailed descriptions of the use cases are provided in Appendix~\ref{sec:appendix-use_case}.

\hyperref[sec:uc:sensorqc]{\emph{Sensor QC}} applies an interpretable graph convolutional network to anomaly detection in irregular environmental sensor networks, emphasizing documented preprocessing and the quality and provenance of expert-generated labels \citep{lasota2025interpretable}. \hyperref[sec:uc:swinpl]{\emph{SWIN-PL}} compares weather-forecasting models using 13 and 37 pressure levels, illustrating the dependence of large-scale training on complete gridded fields, machine-readable metadata, consistent preprocessing, and efficient data access. \hyperref[sec:uc:jSDM]{\emph{jSDM}} jointly models the response of a marine protist community to environmental and spatial drivers, making covariate completeness, bias documentation, and standardized sample metadata particularly important. \hyperref[sec:uc:sinkhole]{\emph{Sinkhole}} instance segmentation uses drone and satellite images to identify individual sinkholes on the shore of the eastern Dead Sea \citep{alrabayah2024}, illustrating the importance of expert annotations, sensor resolution and ground truth target preparation. \hyperref[sec:uc:aqqua]{\emph{AqQua}} is a Helmholtz Foundation Model Initiative (HFMI) project developing a foundation model for marine plankton images. The \hyperref[sec:uc:groundwater]{\emph{GroundwaterMonitoring}} use case is represented twice in the table as there are two consecutive studies with evolving requirements. In \emph{EU}, neural networks and transfer learning are applied to estimate water table depth anomalies, where consistent observational datasets are lacking. In \emph{cUS}, random forests are  trained on observations to estimate water table depth.

\begin{table}[t]
  \centering
  \scriptsize
  \setlength{\tabcolsep}{2pt}
  \renewcommand{\arraystretch}{1.3}
  \caption{Importance of the AI-readiness criteria for the selected use cases
    (columns grouped by category; cf.\ Table~\ref{tab:air-overview}).
    \rfull:~essential, \rmost:~important, \rpart:~nice to have,
    \rwk:~irrelevant; blank:~not applicable.}
  \label{tab:air-usecases}

  \begin{tabular*}{\linewidth}{@{\extracolsep{\fill}}l *{25}{c}@{}}
    \toprule
    \multirow{2}{*}{\textbf{Use case}}
      & \multicolumn{6}{c}{\textbf{Data Preparation (P)}}
      & \multicolumn{4}{c}{\textbf{Data Quality (Q)}}
      & \multicolumn{7}{c}{\textbf{Data Documentation (D)}}
      & \multicolumn{5}{c}{\textbf{Data Access (A)}}
      & \multicolumn{3}{c}{\textbf{Targets (T)}} \\
    \cmidrule(lr){2-7}
    \cmidrule(lr){8-11}
    \cmidrule(lr){12-18}
    \cmidrule(lr){19-23}
    \cmidrule(lr){24-26}
      & \scriptsize 1.1
      & \scriptsize 1.2
      & \scriptsize 1.3
      & \scriptsize 2.1
      & \scriptsize 2.2
      & \scriptsize 3
      & \scriptsize 1.1
      & \scriptsize 1.2
      & \scriptsize 2
      & \scriptsize 3
      & \scriptsize 1.1
      & \scriptsize 1.2
      & \scriptsize 1.3
      & \scriptsize 2.1
      & \scriptsize 2.2
      & \scriptsize 2.3
      & \scriptsize 3
      & \scriptsize 1.1
      & \scriptsize 1.2
      & \scriptsize 2
      & \scriptsize 3.1
      & \scriptsize 3.2
      & \scriptsize 1.1
      & \scriptsize 1.2
      & \scriptsize 1.3 \\
    \midrule

    \textbf{Sensor QC}
      & \rfull & \rpart & \rmost & \rfull & \rmost & \rfull
      & \rwk & \rwk & \rwk & \rmost
      & \rmost & \rfull & \rmost
      & \rmost & \rmost & \rmost & \rwk
      & \rwk & \rwk & \rwk & \rfull & \rfull
      & \rfull & \rfull & \rfull \\

    \textbf{SWIN-PL}
      & \rfull & \rfull & \rmost & \rfull & \rfull & \rmost
      & \rfull & \rfull & \rmost & \rfull
      & \rfull & \rfull & \rfull & \rfull & \rfull & \rfull & \rfull
      & \rpart & \rfull & \rfull & \rmost & \rfull
      & \rfull & \rfull & \\

    \textbf{jSDM (metabarcoding)}
      & \rfull & \rpart & \rfull & \rfull & \rpart & \rfull
      & \rfull & \rfull & \rpart & \rfull
      & \rfull & \rfull & \rfull & \rpart & \rfull & \rfull & \rfull
      & \rfull & \rmost & \rwk & \rfull & \rpart
      & \rfull & \rfull & \rpart \\

    \textbf{Sinkholes}
      & \rmost & \rpart & \rpart & \rfull & \rmost & \rfull & \rmost & \rfull & \rmost & \rfull & \rmost & \rmost & \rmost & \rpart & \rmost & \rpart & \rmost & \rmost & \rmost & \rmost & \rmost & \rmost & \rmost  & \rfull & \rpart \\

    \textbf{AqQua}
    & \rmost & \rfull & \rfull & \rpart & \rpart & \rpart
    & \rpart & \rmost & \rpart & \rfull
    & \rpart & \rfull & \rfull & \rfull & \rpart & \rfull & \rpart
    & \rfull & \rmost & \rfull & \rfull & \rfull
    & \rmost & \rmost & \rfull\\

    \textbf{GroundwaterMonitoring-EU (2022)}
    & \rpart & \rpart & \rwk & \rmost & \rmost & \rmost
    & \rpart & \rpart & \rwk & \rfull
    & \rpart & \rpart & \rpart & \rpart & \rpart & \rpart & \rmost
    & \rfull & \rwk & \rfull & \rfull & \rmost
    & \rfull & \rfull & \rwk\\

    \textbf{GroundwaterMonitoring-cUS (2026)}
    & \rpart & \rpart & \rwk & \rmost & \rmost & \rmost
    & \rpart & \rpart & \rmost & \rfull
    & \rmost & \rmost & \rmost & \rmost & \rmost & \rmost & \rmost
    & \rfull & \rfull & \rfull & \rfull & \rmost
    & \rfull & \rfull & \rmost\\
        
    \bottomrule
  \end{tabular*}
\end{table}

Several requirements are important across most use cases despite their different data modalities and learning tasks. Information about spatiotemporal data resolution and coverage (Q-3), open data formats (A-3.1) and the open and documented representation of targets (T-1.1, T-1.2) are rated as mostly essential. Documented provenance (P-2.1), human decisions during preparation (P-3), machine-readable and standardized metadata (D-1.2, D-1.3), and efficient ML-compatible formats (A-3.2) are also often important or essential. These commonalities identify a baseline of requirements that supports AI workflows across gridded reanalysis data, environmental sensor time series, and molecular sequence data.

The differences between the profiles are equally informative. Data quality criteria are broadly essential for \emph{SWIN-PL}, \emph{jSDM} and \emph{Sinkholes}, whereas several are irrelevant to \emph{Sensor QC} because irregularities and anomalous observations are themselves part of the learning problem. As \emph{AqQua} relies on self-supervised learning on massive amounts of observation data, information about data preparation and data quality is generally less relevant than in the other cases. For \emph{Sensor QC}, data access concerns and AI permissions are irrelevant as the data are available from the same institution. 
Information about targets is consistently relevant, yet standardization (T-1.3) matters only where targets are externally created labels. It is not applicable to \emph{SWIN-PL}, whose targets are future states of the same continuous physical fields, and only of minor importance for \emph{jSDM}, whose response is the observed community matrix itself.
In the \emph{Sinkholes} case, none of the data documentation and access criteria were essential as the key elements of the workflow were primarily depending on data quality and preparation aspects and the collaborators had been involved in the original data acquisition task, thus they could provide missing context directly. The difference in data documentation needs between the EU and US case of \emph{groundwater monitoring} stems from a learning effect, as adherence to FAIR principles was considered more valuable in the later study.

In conclusion, the ratings should not be aggregated into a single readiness score. A high importance rating may indicate that a requirement is already satisfied and consequently demands little user intervention, as with many ERA5 properties in \emph{SWIN-PL}. Conversely, a criterion may require substantial work without being equally relevant to every application. The table consequently supports the central premise of the framework: AI-readiness is use-case-dependent, while transparency about data properties allows users to determine fitness for purpose.

\section{Discussion}
\label{sec:discussion}

As discussed in section \ref{sec:criteria-discussion}, the criteria form a framework that needs to be manifested through concrete implementation. The framework does not prescribe a particular architecture or implementation strategy, as this will depend on the infrastructure at hand. Instead, the criteria reflect the most prominent user needs, and provide an entry point for further refinement. 

The framework combines AI-oriented criteria with known FAIR criteria, leading to overlap with the FAIR framework and mandating aspects seemingly unspecific to AI. However, this is intentional, to ensure that when our framework is used as a holistic checklist, none of the aspects already covered by FAIR are neglected.

The framework aims to guide the handling of data and the conception of data infrastructure services across key stages of the data life cycle, informing the relevant stakeholders about documentation steps and clarifying the reasoning behind documentation. It showcases what kind of downstream usage data providers and managers can enable by preparing datasets accordingly. For instance, when acquiring data in the field, the framework informs about documentation elements such as human operator decisions where the detail knowledge only exists at this stage. During data processing and synthesis of data products, the framework mandates the documentation of processing steps, specifically of aspects such as gap-filling, de-duplication and outlier treatment. Vice versa, ML developers can use the framework as a tool to communicate transparency needs further upstream, understand practical limits in documentation and curation, and to produce reusable datasets. The framework also raises targets as objects to be curated in their own right, covered by well-defined and documented QA/QC procedures. The criteria list can be used to guide practice beyond broader data governance questions.

As part of implementation, a venue of future effort with large automation and value potential includes the idea of \emph{reactive data infrastructures}. Such infrastructures would react to downstream user needs and requests by AI agents. For instance, in an agentic workflow system, a user request for clarification and patching of discovered data gaps could cause user-specific data processing, according to the user's choice of imputation strategy. In such a scenario, processing steps and reasoning behind them could be re-assessed and re-executed, with adaptations based on documentation following criteria of the data preparation category. Such a shift from static documentation to structured executable elements would enable AI agents to facilitate customization much deeper into the data processing and curation workflow.

Related to this, another area requiring further discussion concerns the meaning of AI-readiness in relation to data access by autonomous agents. The criteria do not explicitly address access through APIs, data services, or interfaces designed for agents. Future versions of the framework may need to consider whether service-level accessibility constitutes an additional dimension of AI-readiness that is distinct from mere human-oriented access. Such developments may become particularly relevant as agent-based scientific workflows and machine-to-machine data exchange become more common.

Targets were relevant in all use case examples, though not always as externally created labels, and not all machine learning methods rely on them.
The uptake of targets as a distinct category may also lead to a further distinction and the development of task-specific modules, for which targets are a first examples.

The criteria are also limited in several aspects. They were motivated and designed against the backdrop of the Earth and environmental research domain and may not be directly applicable to cases in other research areas. The original survey that initiated work on the framework was also severely limited in size and scope, which can only be alleviated to some extent by discussion among the wider working group. The use cases likewise illustrate how the criteria may be interpreted; they do not constitute external validation.

A critical question is when \emph{complete} AI-readiness is achieved. The strict answer to this is ``never'', as no dataset can be made ready for all potential downstream use cases a priori. Completeness can only be answered with respect to the concrete use case at hand, as they define what AI-readiness means within their scope. Still, if all criteria are largely fulfilled, a good compromise of AI-readiness across use cases can be achieved.

Finally, the criteria should be viewed as a starting point for community discussion rather than a definitive standard for immediate adoption by European and international data infrastructures. They were informed by literature review, community feedback, survey results, and discussions with data users, data providers and infrastructure providers. Their primary purpose is to stimulate convergence on a shared understanding of what information should be disclosed to enable informed AI use of scientific datasets. Future work should investigate domain-specific priorities, assess the feasibility of implementing individual criteria within existing infrastructures, and explore mappings to emerging machine-actionable metadata standards. Ultimately, the value of the framework lies not in prescribing a single path to AI-readiness, but in creating a basis for dialog between data providers, infrastructure operators, and AI practitioners. As ML training increasingly leverages older datasets that have not been curated for AI-readiness, questions arise about the effects on the data infrastructure landscape in terms of who provides resources and how costs and responsibilities will be shifted. We do not address the feasibility or cost-benefit ratio of implementing procedures that satisfy the criteria in practice here. We encourage all stakeholders in research data management to engage in a discussion from their point of view.

\conclusions  %% \conclusions[modified heading if necessary]

In this article, we present a framework for assessing and improving AI-readiness in research data management practice. The framework extends prior work, most notably the FAIR principles and the ESIP data AI-readiness checklist. It projects ML user requirements into a set of criteria that cover the most essential elements key to downstream use of data, making properties stemming from data generation, capture and processing steps more transparent. It includes a dedicated section on target data for supervised learning applications, motivating that documentation, quality-control and curation procedures should also be extended to instance labeling tasks. The framework should be seen as the start of a discussion and needs further reflection by all stakeholders involved in research data management.

Beyond further discussion among a wider audience and potential transfer to other research domains, future work should focus on the path into practice. The framework can be used for self-assessment, for which tooling could be developed comparable to the ESIP work, and it can also inform independent certification, for which procedures would need to be established. The framework is implementation-agnostic and does not prescribe a particular architecture or implementation path. Additional effort is required by data infrastructures that aim to adopt it. For a concrete data infrastructure, an adoption strategy would need to be developed and tested through practical data workflows, including the implementation or adaptation of relevant services and procedures.

%% The following commands are for the statements about the availability of data sets and/or software code corresponding to the manuscript.
%% It is strongly recommended to make use of these sections in case data sets and/or software code have been part of your research the article is based on.

%\codeavailability{TEXT} %% use this section when having only software code available

%\dataavailability{TEXT} %% use this section when having only data sets available

%\codedataavailability{TEXT} %% use this section when having data sets and software code available

%\sampleavailability{TEXT} %% use this section when having geoscientific samples available

%\videosupplement{TEXT} %% use this section when having video supplements available

\appendix
\newpage

%\begin{landscape}
\setlength{\LTcapwidth}{\linewidth}
\section{Detailed AI-readiness Criteria}
\footnotesize
\renewcommand{\arraystretch}{1.1}
\setlength{\tabcolsep}{4pt}
 
% Column weights sum to 5 (the number of X columns)
\newcolumntype{C}{>{\hsize=0.80\hsize\RaggedRight\arraybackslash}X} % Criterion
\newcolumntype{Q}{>{\hsize=0.85\hsize\RaggedRight\arraybackslash}X} % Question
\newcolumntype{P}{>{\hsize=1.10\hsize\RaggedRight\arraybackslash}X} % Purpose
\newcolumntype{U}{>{\hsize=0.80\hsize\RaggedRight\arraybackslash}X} % Users
\newcolumntype{M}{>{\hsize=1.45\hsize\RaggedRight\arraybackslash}X} % Methods
 
\begin{xltabular}{\linewidth}{@{} C Q P U M @{}}
\caption{Detailed information for each AI-readiness criterion: the
  question it answers, why it matters for AI-readiness, its significance
  for users, and (where possible) illustrative methods. The prefix of each
  AIR-ID denotes the category: data preparation (P), data quality (Q),
  data documentation (D), data access (A), and targets (T). Listed methods
  are illustrative and do not form an exhaustive checklist.}
\label{tab:air-detail}\\
\toprule
\textbf{Criterion} & \textbf{Question} & \textbf{Purpose} & \textbf{Users} & \textbf{Methods} \\
\midrule
\endfirsthead
\multicolumn{5}{l}{\footnotesize\itshape Table~\thetable\ (continued)}\\[2pt]
\toprule
\textbf{Criterion} & \textbf{Question} & \textbf{Purpose} & \textbf{Users} & \textbf{Methods} \\
\midrule
\endhead
\midrule
\multicolumn{5}{r}{\footnotesize\itshape continued on next page}\\
\endfoot
\bottomrule
\endlastfoot
 
% ---------- P ----------
\textbf{P-1.1}\newline Missing values and data gaps have been quantified, documented, and either imputed with declared methods or explicitly left as missing.
& Can we trust that missing data do not systematically distort model outcomes?
& \textit{Prevents hidden bias and ensures model validity.} AI models are highly sensitive to hidden biases introduced by missing data (e.g.\ if data are missing systematically as sensors fail under extreme conditions). Undocumented imputation can distort distributions and lead to unreliable predictions. Moreover, missing points encoded via dummy values may not be obvious to users, leading to biases.
& AI model trainers can reduce effort in independently auditing this data. Model users are subject to less risk in result interpretation and decision-making.
& Missingness is quantified (e.g.\ \% per feature, spatial/temporal gaps) and reported; conditions for systematic missingness are documented. Handling is documented: imputation method (e.g.\ interpolation, model-based) and exclusion thresholds. Remaining gaps are explicitly flagged, not silently filled. Flags are consistent (no mixed use of ``null'', ``N/A'' and dummy values). The impact of imputation on demonstrator model performance is assessed and recorded. \\
\midrule
 
\textbf{P-1.2}\newline Outliers and corrupt values are detected, quantified, and documented.
& Can we trust that outliers or corrupt values do not systematically distort model outcomes?
& \textit{Prevents models from optimizing towards singularities of the data distribution or failing to find optimal solutions due to irrelevant noise.} AI models are sensitive to overfitting and can be confused by data points not representative of the general distribution.
& AI model trainers can reduce effort in independently auditing this data. Model users are subject to less risk in result interpretation and decision-making.
& Statistical or ML-based detection methods are applied, and all identified outliers are logged with decisions on retention, transformation, or removal in preprocessing records. Defining and following adequate operating protocols for observations can ensure faulty measurements are detected and documented. \\
\midrule
 
\textbf{P-1.3}\newline Duplicate data are flagged or removed and the process is documented.
& Do duplicate entries create bias or artificially inflate patterns in the data?
& Duplicates can skew distributions and lead to overfitting or misleading conclusions.
& Data engineers and scientists benefit from cleaner datasets, while decision-makers gain confidence that insights are not artificially amplified.
& The origins of data aggregated from multiple sources are flagged. Exact and fuzzy matching is applied, and deduplication steps are documented and version-controlled. What qualifies as a duplicate should be considered narrowly: observations of the same phenomenon with almost the same value provide a measure of variety and are not usually duplicates. True duplicates may arise when erroneously concatenating overlapping data sequences; careful documentation is required. \\
\midrule
 
\textbf{P-2.1}\newline The provenance of processing steps, tools and parameters is documented.
& Can the origin and transformation history of the data be reliably traced?
& Provenance ensures reproducibility, accountability, and trust in both data and derived models.
& Researchers and auditors use provenance to validate results, while engineers depend on it for pipeline maintenance. Analysis of provenance records can reveal sources of error or bias during ML development cycles.
& Lineage tracking systems and version control record sources and transformation workflows. Examples: origin of the data (e.g.\ sensor type, or source datasets for an aggregation) is recorded; gridding/regularization/resampling is documented including tools and parameter settings; aggregation/smoothing is documented likewise; a potential down-sampling process is documented. \\
\midrule
 
\textbf{P-2.2}\newline The level of data processing is explicitly defined and documented, including distinctions between raw and derived data.
& Can users correctly interpret the processing level and transformations applied?
& Misunderstanding processing levels can lead to incorrect analyses. This criterion also covers provenance, but views the data from its current state rather than its history.
& Analysts and data users depend on this to assess suitability and avoid misuse.
& Metadata includes the timeliness of the data (e.g.\ near-realtime, date of acquisition, date of collation, \ldots); links to published data quality procedures or assessment reports; whether the data is measured or simulated; and any data peer-review process. Processing stages are documented in metadata and supported by lineage records. Optimally, processing scripts are published. \\
\midrule
 
\textbf{P-3}\newline Human operator decisions during data preparation are documented.
& Do undocumented human decisions introduce hidden bias or irreproducibility?
& Manual choices can significantly influence outcomes and must be transparent for reproducibility and auditability. Rich documentation of decisions helps explain deviations.
& Auditors and stakeholders gain trust through transparency, while data scientists can replicate workflows.
& Decision logs and experiment-tracking systems capture manual steps and their rationale. For example, experiments on ships rescheduled due to weather deviate from standard protocol (measurements at a different time of day). In coastal areas, tide-dependent measurements make it impossible to sample at the same time and place. \\
\midrule
 
% ---------- Q ----------
\textbf{Q-1.1}\newline The data include all relevant available data points, and any excluded data are documented along with the reasons for their exclusion.
& \textit{Does the dataset cover the intended domain without critical omissions?} Both Q-1.1 and P-1 concern completeness: P-1 covers data excluded/transformed because they are wrong or faulty, whereas this criterion makes transparent that all relevant acquired data are actually included.
& Omitting data on intransparent criteria can introduce bias and reduce representativeness.
& Researchers and decision-makers depend on completeness to ensure reliable conclusions.
& Coverage metrics are calculated and gaps are documented, along with mitigation strategies where applicable. \\
\midrule
 
\textbf{Q-1.2}\newline Sources of bias are identified, assessed, and documented, along with potential mitigation strategies.
& Do systematic biases affect fairness or validity of outcomes?
& Bias can lead to unfair or incorrect predictions, undermining trust and compliance.
& End users benefit from reduced biases, while auditors require bias documentation for evaluation. ML developers can better understand and mitigate remaining biases.
& Bias analyses are conducted, and mitigation techniques are applied and documented. This applies to the full processing workflow: e.g.\ pre-split training and test sets may inherit or mitigate biases, and these decisions must be documented. \\
\midrule
 
\textbf{Q-2}\newline The metadata contains quantitative measures of uncertainty.
& How reliable are the data and derived results given inherent uncertainty?
& Uncertainty quantification prevents overconfidence and supports informed decision-making. Some ML algorithms exploit uncertainty values to better represent real-world distributions.
& Decision-makers assess risks more effectively and scientists ensure analytical accuracy. ML developers benefit from increased information content.
& Examples: data metrics such as signal-to-noise, (auto-)correlation, uniformity; distribution statistics of a known true distribution from which the data is drawn; confidence intervals and uncertainty metrics included in documentation; a data variable indicating measurement uncertainty per data point. \\
\midrule
 
\textbf{Q-3}\newline The metadata contains quantitative information of data resolution and extent in space and time.
& Are the dataset resolution and extent appropriate for the intended analytical use?
& Resolution determines applicability and prevents misuse at inappropriate scales. Extent helps validate whether the data are fit for purpose.
& Researchers and model developers use this to assess fitness for purpose or when merging datasets of different resolutions.
& Resolution metadata is recorded using standard units and formats (e.g.\ meters per pixel or seconds per integration bin). Spatial and temporal extent is documented using units such as square-kilometers or hours. \\
\midrule
 
% ---------- D ----------
\textbf{D-1.1}\newline The data are accompanied by rich metadata.
& Is sufficient contextual information available to interpret and reuse the data for AI workflows?
& Interprets FAIR principles F2 and R1 for AI workflows. Rich metadata ensures interpretability and reuse when ingesting data into AI workflows or interpreting model results.
& Researchers and data users benefit from improved usability and understanding.
& Assembling rich metadata is complex; we refer to FAIR procedures and give illustrative examples: dataset extent (e.g.\ volume, dimensionality, number of assets, table dimensions) --- relevant for loading scripts and model setup; contact information --- error-mode analysis often needs advice from data experts; timeliness of the data --- relevant when combining sources; links to published quality procedures --- relevant when analysing model error modes. \\
\midrule
 
\textbf{D-1.2}\newline The metadata is machine-readable.
& Can metadata be processed directly by machines?
& Machine-readable metadata supports automation and scalability. ML data libraries often handle standards-compliant metadata out-of-the-box, lessening development effort.
& Developers and AI systems benefit from efficient metadata use.
& A metadata record can be obtained by scripts, e.g.\ by reading file headers or resolving unique identifiers; accessed via a standardized protocol (e.g.\ HTTPS); and expressed in structured formats such as JSON, XML, NetCDF or RDF. \\
\midrule
 
\textbf{D-1.3}\newline The metadata is expressed in compliance with a machine-actionable community standard.
& Is metadata structured to support cross-system integration and compatible with automated workflows and shared ecosystems?
& Standardization enables interoperability and efficient data exchange; compliance ensures consistent integration and reuse across platforms.
& Data engineers and developers benefit from reduced integration effort.
& Metadata should follow best practices (machine-accessibility, availability in perpetuity, persistent identifiers). This criterion further mandates compliance and options for a community-supported ontology or vocabulary. Such resolvable semantic mappings should use standardized persistent links, e.g.\ DOIs, ORCIDs, Handles, RORs. \\
\midrule
 
\textbf{D-2.1}\newline The data dictionary is machine-readable and encoded using an open format.
& Can parameter and variable definitions be accessed programmatically?
& Machine readability enables automation and reduces manual effort. For AI applications, common Python frameworks can load and interpret the dictionary out-of-the-box.
& Developers and AI systems benefit from automated interpretation.
& Formats such as netCDF that support structured fields are used. A counterexample is an instrument vendor-specific format that does not adhere to open industry standards. \\
\midrule
 
\textbf{D-2.2}\newline Dataset variables and parameters are described by a comprehensive data dictionary.
& Are variables clearly defined to prevent misinterpretation?
& Clear definitions ensure correct data usage and interpretation.
& Data scientists rely on accurate feature understanding.
& Data fields (units, variable names, descriptions) are defined. Data structure is documented (which variable is in which column and what unit is used). A structured dictionary is created and maintained alongside the dataset. Examples: NetCDF headers or the NASA-Ames format. \\
\midrule
 
\textbf{D-2.3}\newline Data dictionary entries follow standardized vocabularies and schemas.
& Is variable documentation consistent across datasets?
& Standardization enables integration and comparability.
& Researchers and integrators benefit from harmonized definitions.
& Controlled vocabularies and schemas are applied. Example: NetCDF following the CF-Conventions. \\
\midrule
 
\textbf{D-3}\newline The loading scripts are available and documented.
& Can the dataset be easily reused?
& Script availability supports reusability and transparency.
& Researchers and engineers can set up workflows efficiently; usage barriers are lowered and sources of error reduced.
& Persistent links to code repositories and to interactive tools (e.g.\ notebooks) are given. Ideally, data loaders compatible with standard ML libraries are provided. Example uses are facilitated by interactive tools and example data. Scripts are version-controlled and accompanied by execution instructions. \\
\midrule
 
% ---------- A ----------
\textbf{A-1.1}\newline The data are accompanied by a clear data usage license.
& Are legal usage conditions transparent and unambiguous?
& Clear licensing reduces legal risk and misuse.
& Organizations and users rely on this for compliant data usage.
& Licensing terms are documented, linked to the data, and a link is provided in the metadata. \\
\midrule
 
\textbf{A-1.2}\newline The data usage license is standardized and machine-readable.
& Can licensing terms be automatically interpreted?
& Standardization enables automated compliance checks.
& Platforms and developers benefit from simplified license handling.
& Standard licenses are encoded in machine-readable formats, e.g.\ following the SPDX specification (\url{https://spdx.org/licenses/}). A problem arises if the identifier contains spaces (``CC - BY''), which disturbs machine readability. \\
\midrule
 
\textbf{A-2}\newline The metadata provides information on the consent of data creators to allow AI-based exploitation.
& Is data use with AI aligned with creator consent?
& Proper consent ensures ethical and legal compliance.
& Organizations and ethics boards ensure responsible AI usage.
& Consent records are stored and associated with usage conditions. Ideally, explicit references into the legal frameworks are given. \\
\midrule
 
\textbf{A-3.1}\newline The data are available in at least one open, non-proprietary file format.
& Is the dataset accessible without proprietary tools?
& Open formats ensure accessibility and long-term usability.
& Researchers benefit from avoiding proprietary tool lock-in.
& Data is stored and distributed in open standard formats. \\
\midrule
 
\textbf{A-3.2}\newline The data file format affords ease of use and efficiency in ML workflows.
& Is a format used that is supported by typical ML data libraries? Can data be efficiently loaded?
& Machine-readable formats enable efficient workflows. Widely used ML libraries support loading from standardized formats; efficient loading significantly reduces development effort and improves usability.
& Developers and engineers reduce preprocessing overhead.
& Structured formats such as Zarr, HDF5 or netCDF are used and validated. Larger datasets aggravate problems as data may exceed local or GPU RAM limits; chunking for large multidimensional datasets is essential, with chunk sizes designed to facilitate typical access patterns. \\
\midrule
 
% ---------- T ----------
\textbf{T-1.1}\newline If targets are available, they are provided in an open format.
& Are prediction targets/labels for supervised learning clearly defined and interpretable? Can they be read with non-proprietary software?
& The skill of supervised ML models crucially depends on the extent and accuracy of targets. Targets can be, e.g.\ text labels for classification (``cat'', ``dog''), or target variables for prediction.
& Data scientists rely on this for accurate model development and evaluation.
& Not every dataset has targets, and generating them is not the data provider's task. If targets exist, they follow the same data-management procedures, and the relation between data and targets is encoded (e.g.\ a published compound dataset). Targets are provided in a readable, non-proprietary format. \\
\midrule
 
\textbf{T-1.2}\newline The targets were subject to a documented QA/QC process and the process for generating targets is documented.
& Are targets accurate and consistent? Is the process by which they were created documented?
& Poor target quality degrades model performance. Labels must be consistent (no slightly different labels for the same entity). Generation details (how many annotators, their background, conditions) are relevant for analysing model faults.
& Developers benefit from reliable targets and documented procedures for robust models.
& Which tool was used for manual labeling? Who labeled and how? If labels accompany data as a compound dataset, this is part of its provenance record. Multi-session/annotator merging is consistent. Pitfalls: arbitrary or only-easy annotation; unclear bounding-box rules (e.g. annotate a starfish's buried arms); cryptic, nonsensical or offensive custom labels; class-balance representativeness. \\
\midrule
 
\textbf{T-1.3}\newline The labels follow a controlled vocabulary or ontology.
& Can annotations be reused and integrated across systems?
& Consistent labels maximize interoperability and reuse.
& Researchers benefit from shareable and reusable labels.
& Ideally, the controlled vocabulary is standardized across a domain. The annotation process is documented and quality-controlled, with clear rules on when and how to define label names. \\
 
\end{xltabular}
%\end{landscape}

\section{Checklist to Examine AI-readiness for Open Environmental Datasets}
\label{sec:appendix-esip}
Many of our criteria are closely related to the Checklist to Examine AI-readiness for Open Environmental Datasets \citep{esip2023checklist}. The original checklist did not contain good reference numbers; therefore, to better indicate the connections, we introduce specific identification numbers for each of the criteria which are specified on the following page.

\begingroup
\footnotesize
\setlength{\tabcolsep}{4pt}
\renewcommand{\arraystretch}{1.1}
\noindent
%---------------- LEFT BLOCK ----------------
\begin{minipage}[t]{0.49\textwidth}
\begin{tabular}{@{}p{0.9cm}p{5.6cm}@{}}
\multicolumn{2}{@{}l}{\textbf{1\quad Data Preparation}}\\
\midrule
1.1 & Have null values/gaps been filled? \\
1.2 & Have outliers been identified? \\
1.3 & Is the data single-source or aggregated from several sources? \\
1.4 & Has the data been gridded (regularized in space and time) or is it in the originally sampled resolution? \\
1.5 & Have targets been identified and labeled (i.e.\ usable as a supervised learning training dataset)? \\
\addlinespace[4pt]
\multicolumn{2}{@{}l}{\textbf{2\quad Data Quality}}\\
\midrule
2.1 & Have measures been taken to ensure completeness? \\
2.2 & Are there automated processes to monitor consistency? \\
2.3 & Have measures been taken to reduce bias? \\
2.4 & What is the timeliness of the data? \\
2.4.1 & Difference between raw near-real-time access vs fully quality-controlled data with additional delay? \\
2.5 & Are there quantitative measures of uncertainty? \\
2.6 & Is there quantitative information about data resolution in space and time? \\
2.7 & Are there published data quality procedures or reports? \\
2.8 & Is the provenance tracked and documented? \\
2.9 & Are there checksums / other checks for data integrity? \\
2.10 & How big is the dataset? \\
2.11 & Is this raw data or a derived/processed data product? \\
2.12 & Is this observational data or simulation/model output? \\
2.13 & Has the data been peer-reviewed? \\
2.14 & Has it been down-sampled to reduce resolution, or is it raw? If so, are the raw data available? \\
\multicolumn{2}{@{}l}{\textbf{3\quad Data Documentation}}\\
\midrule
3.1 & Does the dataset have metadata? \\
3.1.1 & Is the dataset metadata standardized? \\
3.1.2 & Is the dataset metadata machine-readable? \\
3.1.3 & Does it include details on the spatial and temporal extent? \\
\end{tabular}
\end{minipage}
\hfill
%---------------- RIGHT BLOCK ----------------
\begin{minipage}[t]{0.49\textwidth}
\begin{tabular}{@{}p{0.9cm}p{5.6cm}@{}}
3.2 & Is there a comprehensive data dictionary/codebook to describe parameters? \\
3.2.1 & Is the data dictionary standardized? \\
3.2.2 & Is the data dictionary machine-readable? \\
3.2.3 & Do the parameters follow a defined standard? \\
3.2.4 & Are parameters crosswalked in an ontology or common vocabulary (e.g.\ NIEM)? \\
3.3 & Does the dataset have a unique persistent identifier, e.g.\ DOI? \\
3.4 & Is there contact information for subject-matter experts? \\
3.5 & Is there a mechanism for user feedback and suggestions? \\
3.6 & Are there example codes / notebooks / toolkits showing how the data can be used? \\
3.7 & Is there a clear data license? \\
3.7.1 & Is the license standardized and machine-readable (e.g.\ Creative Commons)? \\
3.8 & Has this dataset already been used in AI or ML activities? \\
3.9 & Are there recommendations on intended (and non-recommended) use of the data? \\
\addlinespace[4pt]
\multicolumn{2}{@{}l}{\textbf{4\quad Data Access}}\\
\midrule
4.1 & What is the file format? \\
4.1.1 & Is it machine-readable? \\
4.1.2 & Is it available in at least one open, non-proprietary format? \\
4.1.3 & Is it available in several different file formats? \\
4.2 & Data delivery \\
4.2.1 & Direct file download or ordering? \\
4.2.2 & Is there an API? \\
4.2.2.1 & Custom-developed or open, standard protocol? \\
4.2.3 & On the cloud? \\
4.3 & For restricted data, are there measures to provide some access while protecting privacy and security? \\
4.3.1 & Has the data been aggregated to reduce granularity? \\
4.3.2 & Has the data been anonymized / de-identified? \\
4.3.3 & Is there secure access to the full dataset for authorized users? \\
\end{tabular}
\end{minipage}
\endgroup

\section{Detailed Use Cases}\label{sec:appendix-use_case}

Below we provide detailed descriptions of the AI-readiness use cases briefly mentioned in the main paper.

\subsection{Sensor QC: Quality Control of Environmental Sensor Data}
\label{sec:uc:sensorqc}

The goal of this study was to automate quality control (QC), with a focus on anomaly detection, in environmental sensor networks with sparse, irregular spatial layouts while ensuring interpretability of results \citep{lasota2025interpretable}.

We developed a Graph Convolutional Network (GCN) that models the irregular sensor network as a graph structure. The GCN outperforms the graph-less baseline Long Short-Term Memory (LSTM) network which did not incorporate information from adjacent sensors. To ensure interpretability, integrated gradients were used as an Explainable AI (XAI) method to map how the time series data of individual neighboring stations influence an anomaly classification.

The most prominent challenges related to data were the highly heterogeneous quality of the data and the tedious ground-truth labeling required for the supervised learning that relied heavily on manual, and thus somehow subjective, expert annotation.

\subsubsection{Data source(s)}

Data from two different environmental sensing networks was used:  1. Attenuation data from Commercial Microwave Links (CMLs), an opportunistic data source available from the backhaul of the mobile phone network. 2. In situ soil moisture and temperature measurements managed from the SoilNet network operated by Helmholtz Centre for Environmental Research (UFZ) within TERENO.

\subsubsection{AI-readiness evaluation}

For this use case, the most relevant AI-readiness criteria focus on data preparation, specifically P-1.1, P-2.1, and P-3. While our anomaly detection model aims for maximum robustness, its performance remains sensitive to strong variations in preprocessing and labeling procedures. Ensuring consistency and documentation in these initial stages is therefore essential for reproducible results.

\subsection{SWIN-PL: Vertical-Resolution Study on Weather Forecasting}
\label{sec:uc:swinpl}

This use case investigates whether the skill of a data-driven global weather forecasting model benefits from increasing the vertical resolution of its atmospheric input and output fields from 13 to 37 pressure levels. In particular, the experiment separates two possible benefits of the additional levels: finer sampling within the atmospheric range already represented by the 13-level model, and greater vertical reach into the upper atmosphere.

The forecasting system follows a Pangu-style shifted-window Transformer architecture. Two consecutive atmospheric states are passed through an input head that embeds the gridded fields into patches, a SWIN U-Net core that encodes and decodes the spatial representation, and an output head that recovers the forecast fields. The separation between input-/output-heads and the core enables a frozen-core experiment in which a core trained with 13 pressure levels is reused with newly trained 37-level input and output heads. Comparing this adapted model with a model trained natively on all 37 levels tests whether the 13-level core has already learned representations that support predictions at additional levels, or whether the full model must be retrained to acquire the necessary information.

\subsubsection{Data source(s)} The data source is the ERA5 reanalysis \citep{hersbach2020era5}, accessed through WeatherBench~2 \citep{rasp2024weatherbench2} in a cloud-optimized Zarr format. The dataset contains upper-air variables on pressure levels and surface variables on a regular global grid over a multi-decadal period.

\subsubsection{AI-readiness evaluation}
Because ERA5 is a mature, heavily curated reanalysis, it requires little active intervention from the user. However, this high degree of out-of-the-box readiness should not be conflated with low importance. On the contrary, high AI-readiness across almost all dimensions was a prerequisite for this project, as training high-capacity SWIN Transformers fundamentally relies on massive, well-structured, and machine-actionable datasets.

\emph{Data Preparation (P).}
The near-gap-free atmospheric fields were critical, as data imputation at this scale would be prohibitive. Two preparation steps nevertheless required significant effort. First, the pressure-level fields had to be restacked so that each variable-level pair became an individual input channel, collapsing the Zarr \texttt{level} dimension into the channel axis. Second, sea-surface temperature, which ERA5 stores as \texttt{NaN} over land, was replaced with a constant value of \texttt{0}. Because ERA5 represents temperature in Kelvin, this value lies outside the physical range of SST, providing the network with an unambiguous land sentinel while avoiding \texttt{NaN}s that could break training.

\emph{Data Quality (Q).}
High data quality was essential for this project. Complete spatial coverage, precise high-resolution grids, and temporal consistency are mandatory constraints for training global autoregressive models. Uncertainty quantification required less active management: known reanalysis biases were inherited from upstream rather than reassessed here, and the deterministic WeatherBench subset provides no per-grid-point uncertainty.

\emph{Data Documentation (D).}
Working with 37 vertical levels across multiple atmospheric variables makes strict documentation and metadata essential. The project relies on standard climate-variable naming conventions and on machine-readable Zarr metadata and data dictionaries. Without these, programmatic loading and the mapping of variables and levels to model channels would be highly error-prone.

\emph{Data Access (A).}
Given the volume of high-resolution, 37-level ERA5 data spanning decades, machine-optimized access was critical. Open availability in a cloud-optimized Zarr format was a key enabler, allowing high-throughput training directly from cloud storage. The lack of a machine-readable license identifier was the principal remaining limitation.

\emph{Targets (T).}
The targets are future ERA5 states used for autoregressive forecasting and therefore inherit the same requirements for quality, access, and documentation as the inputs. Their open representation and correct temporal alignment remain essential. Because they are derived directly from the dataset rather than supplied as external semantic labels, a controlled target vocabulary is not applicable.

% ############## Bio Use Case ##############################
\subsection{Joint Modeling of Protist Community Composition at HAUSGARTEN}
\label{sec:uc:jSDM}
This use case quantifies the drivers of decadal change in marine protist community composition. The data originate from the multi-year sampling series at the HAUSGARTEN long-term ecological research (LTER) observatory in the Fram Strait \citep{soltwedel2005}. The model estimates how much of the observed compositional variation each of three sources explains: measured local environmental conditions, spatial and temporal structure, and residual covariation among taxa that the covariates leave unexplained. This partition matters for interpreting community reorganization under Atlantification. A shift toward boreal taxa has two possible causes. Either the local environment has changed and the community follows, or the taxa arrive from elsewhere, carried in by Atlantic water. The model separates these cases: it shows whether the measured environmental covariates explain the shift or not. 
Joint species distribution models (jSDMs) model all taxa at once instead of fitting one regression per taxon. Each taxon keeps its own response to the environmental covariates through a generalised linear predictor. A residual covariance term captures the correlations among taxa that remain after the model accounts for the environmental and spatial components. A full covariance matrix over thousands of taxa is intractable, so implementations approximate it with a few latent factors. These factors also let rare taxa borrow strength from taxa with similar responses. Scalable implementations handle thousands of taxa and can replace the linear predictor with a neural network for non-linear responses \citep{pichler2021}. The model partitions the variance into environmental, spatial, and residual association components, and the association matrix generates hypotheses about ecological coupling between taxa. 
Two features shape what this case needs for AI-readiness. First, it reverses the usual roles: the amplicon sequence variant (ASV) table is the response, and the environmental metadata is the predictor set. AI-readiness therefore hinges on the completeness and comparability of the covariates, which are often curated less carefully than the sequence data. Second, the output is inferential, not predictive, so held-out performance cannot catch a spurious association or a misattributed variance component. Here, AI-readiness depends on transparent documentation of the upstream preparation decisions.

\subsubsection{Data source(s)}
The data for metabarcoding studies come from several distinct sources. 
The primary source is the raw sequencing output: paired-end Illumina reads of a marker region, here the V4 region of the 18S ribosomal RNA (rRNA) gene, which is a widely  used barcode for eukaryotic and, in particular, protist diversity \citep{stoeck2010}. These reads are deposited in a public sequence archive such as the European Nucleotide Archive (ENA) at the European Bioinformatics Institute \citep{ena2021}, whose holdings are mirrored daily across the International Nucleotide Sequence Database ollaboration (INSDC) \citep{arita2021}, so a submission to one archive is available through all of them. 
The second source is the processed sequence data. A chain of processing steps turns the raw reads into an amplicon sequence variant (ASV) table, which records how often each exact sequence variant occurs in each sample: quality filtering and trimming, denoising into exact variants, merging of paired reads, and chimera removal \citep{callahan2016}. Post-clustering curation then merges variants that co-occurrence patterns identify as artefacts of the same biological entity \citep{froslev2017}. 
Where the analysis assigns taxonomy to these variants, it draws on versioned reference databases such as the Protist Ribosomal Reference database (PR2; \citealp{guillou2013}) or SILVA \citep{quast2013}, and the specific release is recorded.
A third source is the environmental context of each sample, which supplies the covariates. These observations come mainly from conductivity-temperature-depth (CTD) casts together with the associated nutrient and pigment measurements, and Earth and environmental projects archive them in repositories such as PANGAEA \citep{felden2023}. Gridded remote-sensing products and reanalyses, for example sea-ice concentration and sea-surface temperature from the ERA5 reanalysis \citep{hersbach2020era5}, can extend this set. Sample provenance, cruise, depth, collection date, coordinates, and sequencing protocol follow the community standards for minimum information about a marker gene sequence and about any sequence (MIMARKS and MIxS; \citealp{yilmaz2011}), which make the join between the sequence and environmental sources reliable.

\subsubsection{AI-readiness evaluation}

\emph{Data Preparation (P).} Missing covariate values (P-1.1) are the most consequential issue, because a sample with an incomplete predictor vector is dropped from the model. In polar fieldwork these gaps are non-random: casts abandoned in ice or bad weather remove the extreme conditions that are most informative about environmental response, so undocumented listwise deletion biases the fitted response curves. Deduplication (P-1.3) also demands care. Sequence variants that differ by a few bases may be biological or technical in origin, and uncollapsed variants enter the model as near-perfectly correlated taxa, which produces residual associations that reflect pipeline artefacts, not ecology. The thresholds and parameters of denoising, curation, and any prevalence filtering are therefore modelling-relevant and must be logged, not assumed (P-2.1), as must operator decisions such as station relocation or off-protocol sampling times (P-3). Because ASV counts are compositional and depend on sequencing depth \citep{gloor2017}, the transformation that turns the table into model features is itself a processing step that must be declared (P-2.2).

\emph{Data Quality (Q).} Bias documentation (Q-1.2) is essential here, with a sharper consequence than in a predictive setting. Technical variation from extraction protocol, PCR conditions, or sequencing batch \citep{sinha2017,esling2015} is often confounded with cruise, and therefore with year. The model splits variance into environmental, spatial, and residual components, so a batch effect is absorbed into one of them and becomes indistinguishable from the ecological trend the study targets. Which taxa the analysis retains is equally a quality matter (Q-1.1): prevalence and abundance filters routinely remove most of the taxon set to make the covariance tractable, and because the discarded tail carries the rare taxa that latent factors are meant to inform, the thresholds and the resulting exclusions must be reported rather than left implicit. Resolution and extent information (Q-3) is equally essential, because station coordinates and sampling dates enter the model directly as its spatial and temporal structure, not as descriptive metadata; expedition-based revisits produce irregular intervals that must be represented explicitly.

\emph{Data Documentation and Access (D, A).} The workflow joins two independently curated data streams, so a comprehensive data dictionary for the environmental covariates (D-2.2), covering units, instrument, and processing level, is as important as the documentation of the sequence data. Rich machine-readable metadata (D-1.1 to D-1.3) is required to perform this join reliably across a decade of cruises. That join also depends on the covariates and the sample descriptors being expressed through standardized vocabularies (D-2.3): the sequence side follows MIxS, while the environmental side inherits whatever terms its repository applies, and the two must be reconciled before the streams can be merged programmatically. Documented loading scripts for the pipeline and for the feature/response join (D-3) reduce the reimplementation effort. Licensing terms (A-1.1) need to be clear on both streams and expressed in an open, non-proprietary format (A-3.1), since sequence reads and environmental observations are deposited under separate repository policies and the license of the joined product is only as permissive as the more restrictive of the two. Access efficiency (A-3.2) matters less here than in the image and reanalysis use cases: a community matrix of this size fits in memory, and the binding constraint is model parameterization, not data throughput. 

\emph{Targets (T).} This use case sits between externally labeled and self-derived targets. The response is the community matrix itself, so no separate annotation step creates it, and its quality assurance is that of the processing pipeline (T-1.2). Open provision of the response matrix and its representative sequences (T-1.1) remains important for reproducibility. A controlled taxonomic vocabulary (T-1.3) helps with interpreting and aggregating results, but the model operates on ASVs, not on taxonomic labels, so annotation errors affect interpretation more than fitting. This case shows that the target category can be relevant without the dataset holding labels in the conventional sense.

% ############## Sinkhole Use Case ##############################

\subsection{Sinkhole: Instance Segmentation in the Eastern Dead Sea}
\label{sec:uc:sinkhole}
\cite{alrabayah2024} developed a U-Net-based workflow to delineate sinkholes at Ghor Al-Haditha on the eastern Dead Sea shore. Training on high-resolution drone imagery was followed by transfer learning to lower-resolution satellite imagery. Initial semantic segmentation distinguished sinkhole and background pixels, with an additional edge pixel class for the drone image case depicting merging sites between two neighboring sinkholes. This ternary segmentation was followed by subsequent post-processing steps to produce individual sinkhole instances within each scene. The overall project illustrates how expert annotation, sensor resolution, and ground truth target preparation influence the suitability of observational imagery for supervised learning.

\subsubsection{Data source(s)}
The inputs were a large December 2016 drone orthophoto scene at 0.1 m/pixel, as well as satellite pan-sharpened Pleiades Neo imagery from August 2022 at 0.3 m/pixel. Sinkhole outlines were traced by hand in ArcGIS, with elevation information supporting the drone annotations. Drone labels included a derived edge / boundary class to distinguish neighboring sinkholes separated by a narrow wall. In satellite imagery, merged groups were labeled as single sinkholes because the in-between boundaries of merged sinkholes were harder to see reliably. TIFF/GeoTIFF imagery and TIFF masks supported a patch-based Python workflow.

\subsubsection{AI-readiness evaluation}
For this particular use case, documenting how training labels are created and checked is essential for assessing the data’s suitability for AI. It is important to know for example that tracing the narrow wall between merged sinkholes required expert judgement and certain assumptions on their part; and that the preprocessing code for instance extracted and widened the boundary of the labels, i.e. thickening them slightly. Sinkholes largely hidden by vegetation were left unlabeled, and those pixels subsequently became examples of the background class, when included in training. These choices make documenting preparation steps, human decisions and label quality checks essential (P-2.1, P-3, T-1.2). Assessing unequal pixel counts across classes, i.e. imbalanced classes, and possible labeling biases is also essential (Q-1.2). Acquisition dates, pixel sizes and the areas covered determine which drone and satellite comparisons are meaningful (Q-3). Numerical estimates of label uncertainty would help describe unclear boundaries (Q-2); the study assigned one class per pixel and proposed probabilistic labels as future work. Sharing source files and code also showed the importance of explaining label meanings and providing documented loading scripts (D-2.2, D-3). Our data consisted of RGB GeoTIFF images and spatially aligned label masks representing a small set of classes, hence this use case did not really require a separate formal data dictionary. Standardized dictionaries generally offer greater value for datasets containing many named variables, such as those often stored in NetCDF, HDF5 or Zarr. Permissions were handled through the supervising researcher who acquired and gathered the overall data. AI-use consent was not recorded in metadata, since LLMs were not mainstream at the time. Today however, we would record the applicable license and consent conditions in machine-readable metadata to clarify permitted reuse (A-1.1, A-1.2, A-2).

% ############# Aqqua Use Case #######################
\subsection{AqQua: A Foundation Model of Plankton Imaging}
\label{sec:uc:aqqua}
Funded under the Helmholtz Foundation Model Initiative (HFMI, https://hfmi.helmholtz.de/), the \emph{AqQua} project (The Aquatic Life Foundation Project, www.aqqua.life) aims to construct the first domain-specific AI foundation model for pelagic imaging. Its primary ambition is to resolve fundamental uncertainties surrounding the marine biological carbon pump, plankton biodiversity, and ecosystem dynamics under climate change by mapping species distributions and calculating carbon export fluxes across unprecedented spatial and temporal scales. To overcome the extreme scarcity of manual labels in marine observation, \emph{AqQua} adapts large-scale self-supervised learning (SSL) paradigms - specifically training vision transformers on high-performance computing infrastructure - to learn generalized representations from raw imagery. Subsequently, these large foundational backbones are distilled into lightweight, cross-platform tools to serve a global community that includes marine ecologists, environmental policymakers, monitoring agencies, and computer vision researchers. Anchored in Open Science principles, \emph{AqQua} will release its trained models, embedding spaces, and exploration interfaces under open licenses, alongside providing public computer vision benchmarks to advance self-supervised methodologies for aquatic data.

\subsubsection{Data source(s)}
The project’s training dataset is assembled from an unprecedented multi-institutional aggregation effort, targeting several billion pelagic images spanning marine and freshwater ecosystems around the world. A substantial core of this repository originates from key Helmholtz Centers—including GEOMAR Helmholtz Centre for Ocean Research Kiel, Helmholtz-Centre Hereon, the Alfred Wegener Institute Helmholtz Centre for Polar and Marine Research (AWI), and the Helmholtz Centre for Environmental Research (UFZ) - and is heavily augmented by voluntary contributions from over 40 global academic laboratories and international partners. The data encompass a vast spectrum of acquisition modalities, ranging from in-situ imaging systems (such as permanently moored camera arrays, autonomous underwater vehicles, and towed instruments such as Underwater Vision Profilers (UVPs) or Virtual Plankton Recorders (VPRs) to laboratory slide scanners, benchtop microscopes, and community-tier devices such as the PlanktoScope, as well as laboratory monocultures. Spatially, the records span coastal shallows, the open ocean, deep-sea layers, and inland lakes, capturing everything from microscopic phytoplankton to macro-zooplankton and particulate organic aggregates. \emph{AqQua}-developed pipelines convert these disparate raw inputs into a unified, high-throughput storage format - combining Blosc2-compressed images inside Lightning Memory-Mapped Databases (LMDB) with unified metadata stored in Parquet files—scheduled for public release under a CC BY-NC 4.0 license by July 2027.

\subsubsection{AI-readiness evaluation}
The importance of AI-readiness criteria for \emph{AqQua} is heavily governed by the unique mechanics of its self-supervised pretraining pipeline and downstream environmental mapping goals. Top-tier priority is assigned to data integrity, open machine-actionability, and spatial grounding: corrupt values and duplicate entries must be programmatically flagged to avoid corrupting the self-supervised embedding space - i.e., preventing train-test-leakage; metadata must adhere to machine-readable community standards with precise quantitative spatiotemporal coordinates (latitude, longitude, depth, time) and pixel scales to enable dimensional anchoring; and open, high-throughput storage formats (LMDB/Parquet) are mandatory to sustain GPU utilization on exascale clusters without I/O bottlenecks. Clear licensing (CC BY-NC 4.0) and explicit contributor consent for AI exploitation are equally critical to secure legal compliance for the planned open-access model and data release. In contrast, some data preparation criteria are deliberately deprioritized to make harmonizing thousands of datasets feasible. Only a defined set of required metadata is retained from the available raw metadata to ensure consistency and quality (with transformations and enrichments applied beyond pure metadata mapping), so some acquisition- or processing-specific metadata (P-2.1) and human operator decisions (P-3) predating the harmonization pipeline may be lost. However, processing steps, tools, parameters, and human decisions within the \emph{AqQua} pipeline itself are documented, and it is made explicit that \emph{AqQua} provides aggregated derived data collection rather than raw acquired data.
Target harmonization poses additional scalability challenges (T-1.2, T-1.3). Because targets are not required for self-supervised pre-training, their absence does not exclude a dataset from inclusion, though they remain essential for fine-tuning, validation, and benchmarking. Since target data vary substantially across datasets and lack a standardized scheme, \emph{AqQua} can only partially retain targets by automatically mapping them to a standard such as \emph{WoRMS} (World register of marine species). Moreover, existing benchmark datasets might not follow such a taxonomy (in conflict with T-1.3), but will still be used for comparison with prior work.

% ############# Groundwater Monitoring Use Case #######################
\subsection{Advancing Groundwater Monitoring}
\label{sec:uc:groundwater}

The use case comprises two studies: one in the EU \citep{ma2022} and one in the contiguous US \citep{ma2026}, both on advancing groundwater monitoring in 2022 and 2026. The methodology shifts from a model-informed, temporal, deep-learning or transfer-learning approach in \cite{ma2021} to an observation-driven, high-resolution, spatial, ensemble-tree approach in \cite{ma2026}. Consequently, the AI-readiness requirements also change.

In 2022, Ma et al. used an LSTM neural network combined with transfer learning (LSTM-TL) to reconstruct monthly water-table depth anomalies across Europe with a resolution of approximately $12.5$ km. The LSTM learns the temporal relationship between precipitation and soil moisture anomalies and groundwater anomalies from the TSMP physically based model. Then, that learned relationship is transferred to observational or reanalysis inputs. This enables the estimation of groundwater anomalies even in areas with sparse groundwater observations. Depending on the evaluation metric, the reconstructed product achieved reasonable agreement with observations from $2,569$ monitoring wells and was subsequently used to reconstruct European groundwater dynamics from the early 1980s onward.

In 2026, a substantially different, more spatially oriented approach is taken: a random forest ensemble with $300$ decision trees is trained directly on more than one million groundwater well observations and high-resolution spatial predictor variables. Rather than modeling temporal groundwater dynamics, the model learns the spatial relationship between long-term mean water-table depth and climate, terrain, and subsurface characteristics. The resulting product has a resolution of approximately $30$ m across the contiguous US.

\subsubsection{Data source(s)}
In 2022, the training data are primarily TSMP-G2A, a continental-scale, physically based coupled atmosphere–land–surface\-groundwater simulation. The LSTM is trained on simulated groundwater-depth anomalies from 1996–2016, using precipitation and soil-moisture anomalies as predictors. For the actual transfer step, these predictors are replaced by corresponding observation-based/reanalysis datasets, allowing the model to generate groundwater anomalies beyond the spatial and temporal coverage of the groundwater observations. The resulting estimates are evaluated against a compiled dataset from $2,569$ European groundwater monitoring wells.

In 2026, the training observations consist of long-term mean water-table-depth measurements from three major US groundwater databases, covering 1914–2023 and yielding $726,828$ grid cells with observations. These are combined with ten spatially gridded climatological and hydrogeological predictors, including precipitation, temperature, potential moisture/evapotranspiration, hydraulic conductivity, soil texture, elevation, slope and distance to streams, at resolutions from approximately 1 arc-second to $1$ km. For the subsequent groundwater-storage calculation, the WTD product is combined with 10-layer porosity data from ParFlow CONUS 2.0.

\subsubsection{AI-readiness evaluation}
In both the 2022 and 2026 studies, the data-access and targets criteria are weighted stronger than the data preparation, quality, and documentation criteria. The availability of clear, AI-oriented license information (A-2) is a major focus. This not only refers to clear terms in data usage, and the ability to load data into a NumPy array or Pandas DataFrame (A-3.1), it also means an LLM can more easily access and read the metadata and therefore find the data. 

The quality of the documentation of the targets (T-1.1 and T-1.2) is usually relevant to the AI model used to generate the data. Ideally, a description of the model and the scripts used to implement it would be provided. The label can be named in any way, as long as it is clearly and thoroughly described. Often, labels are renamed according to user-preferred conventions before training an AI model.

The data preparation criteria are considered less important because missing values, outliers, and duplicates were removed prior to training. However, it is understood that good documentation of these steps is valuable.

A learning effect can also be observed from the 2022 study to the 2026 study. While the FAIR criteria, such as rich metadata or metadata with machine-actionable features (A-1.2, D-1.1 -- D-1.3, D-2.1 -- D-2.3, T-1.3), were considered relevant only for reproducing the team’s 2022 results, these features became much more important by 2026.

\section{Survey}\label{sec:appendix-survey}    %% Appendix A

\subsection{Survey setup}
\label{sec:survey}

In order to better understand and represent requirements from a wide range of users, a digital survey was conducted to let ML users voice their experiences when working with data beyond anecdotal evidence solely from working group members. The survey asked a user to consider a specific case from their own experience working with scientific datasets where they applied ML methods. The survey did not provide a definition of AI-readiness upfront. It requested details on properties of the dataset used and steps conducted when using it for ML; more details are given in appendix section~\ref{sec:appendix-survey-structure}.

The survey was distributed among research groups from the home institutes of the working group members, with the aim to receive responses from colleagues that use or develop ML methods on scientific data. Colleagues who were known to be working with ML methods were specifically encouraged to participate, aiming at particularly detailed answers. The survey was anonymous in the sense that no personal contact details were requested; however, it might be possible for working group members to identify respondents based on their described case. For these reasons, and because no consent for publication was requested, survey results are only available upon request.

A total of 63 raw survey responses were received, only 10 of which could be considered for evaluation as the others were incomplete (the survey tool stored every response even if no fields were actually filled). Feedback from colleagues who did not finish the survey often indicated that they did not have sufficient knowledge of ML methods to answer the survey, possibly indicating that the survey targeted a strong niche of users despite the presumed wide-spread uptake of ML methods in science.

For various reasons, the survey cannot be considered fully representative or unbiased, originating particularly from the way individual respondents were sought and encouraged and due to the limited number of responses. Nevertheless, its results were informing the initial work on criteria and provide additional weighting.

\subsection{Survey structure}\label{sec:appendix-survey-structure}

The survey asked for the following elements:
\begin{itemize}
    \item A brief textual description of the use case, indicating the research question, goal and datasets involved.
    \item Properties of the dataset or datasets:
    \begin{itemize}
        \item storage location (e.g., on local institute servers or cluster; obtained from a public or scientific repository)
        \item whether the data was already published (referenceable via DOI)
        \item total size of the dataset in order of magnitude
        \item use/presence of labels, if applicable (e.g., were labels already included; did the scientist create or add labels)
        \item estimates of what was most helpful and what was most time-consuming when understanding the data (e.g., plotting using own scripts; reading file-level metadata; reading documentation or articles; asking a colleague more familiar with the data)
        \item report on experience with the quality of the data (e.g., how far the dataset was quality-assured; whether issues with the data quality were discovered while working with it; whether the quality of the labels was sufficient)
        \item a rating of overall AI-readiness of the data from the individual perspective of the user
    \end{itemize}
    \item An assessment in the user's own words about:
    \begin{itemize}
        \item What does AI-readiness data mean to them in the context of this specific case?
        \item What would have been useful to have / know about the dataset in view of AI-readiness?
        \item Any other comments.
    \end{itemize}
\end{itemize}

\subsection{Discussion of Survey Results}
\label{sec:survey:discussion}

Several observations could be made from the survey results.

\begin{itemize}
    \item Comments reveal that supervised learning is a common scenario, for which labels of high quality are essential. Most use cases required labels, and in more than half of these cases, some or all of the labels were created by the scientists as the original dataset labels were non-existent or insufficient. It should be noted, however, that labeling is often also use-case dependent and it is impossible to provide labels for a dataset a priori for all its potential downstream uses. Challenges related to labeling included that labels need to be available in a machine-readable standard format, metadata about labels must be made available (e.g., quantity, biases, representativeness, procedures used to generate labels), and the structure of data and labels should be harmonized so they can be used in sync more easily.
    
    \item While both are used, file-level metadata is even more important or directly useful than additional metadata stored beside the files. The reason for this is that self-describing data's metadata (e.g., those stored in NetCDF headers) becomes easily available for further processing, as it is supported and automatically exploited by common software libraries such as \texttt{xarray}. Next to plotting data using own scripts and asking colleagues, this was considered the most useful element in understanding the data.
    
    \item Plotting the data in a multitude of ways is the most time-consuming task. It is probably difficult to ease this by means of research data management processes such as better documentation or viewing services. Ultimately, this is manual, investigative work that is to larger extents unavoidable, but also not seen as a strong burden by users as it helps them to get to know the intricacies of the data at hand.
    
    \item ``Asking a colleague'' is always done and almost always a key element in understanding data. Nevertheless, this does not invalidate the usefulness of other means for understanding data (e.g., metadata, documentation, reference articles) that are recognized and used as well.
    
    \item There is wide variety in the maturity of data used for AI as perceived by the users, e.g. some colleagues used published, quality-assured, mature data, others used data that are unpublished and largely uncurated (gathered in a research project or only available through colleagues). Both extremes exist and are relevant in practice. Even when data are formally published and quality-assured, this does not preclude discovering issues (such as missing units) as data processing and ML model development proceed step-by-step.
\end{itemize}

In summary, the survey offered insights that went into shaping the AI-readiness criteria and can further inform future service development. Despite the issues of the survey regarding its distribution, biases and low number of samples, we currently do not believe that a repetition of the survey with an extended audience would significantly change the gathered insights. The responses showed a good sample of potential cases, and, where applicable, the requested ranking (vote on a scale of 1--5) was fully spread.

%\subsection{}     %% Appendix A1, A2, etc.

\noappendix       %% use this to mark the end of the appendix section. Otherwise the figures might be numbered incorrectly (e.g. 10 instead of 1).

%% Regarding figures and tables in appendices, the following two options are possible depending on your general handling of figures and tables in the manuscript environment:

%% Option 1: If you sorted all figures and tables into the sections of the text, please also sort the appendix figures and appendix tables into the respective appendix sections.
%% They will be correctly named automatically.

%% Option 2: If you put all figures after the reference list, please insert appendix tables and figures after the normal tables and figures.
%% To rename them correctly to A1, A2, etc., please add the following commands in front of them:

\appendixfigures  %% needs to be added in front of appendix figures

\appendixtables   %% needs to be added in front of appendix tables

%% Please add \clearpage between each table and/or figure. Further guidelines on figures and tables can be found below.

\authorcontribution{Conceptualization, methodology, writing - original draft, writing - review and editing - all authors.} %% this section is mandatory

\competinginterests{SKH is a member of the editorial board of \emph{Earth System Science Data}.} %% this section is mandatory even if you declare that no competing interests are present

%\disclaimer{TEXT} %% optional section

\begin{acknowledgements}
We would like to thank the anonymous survey contributors for their input. AI tools were used in the preparation of this manuscript to refine the academic tone and stylistic flow of this paper. All final content was verified by the human authors, who remain fully accountable for the published work.
TW and DC were supported by Helmholtz Association's Initiative and Networking Fund through Helmholtz AI [grant number: ZT-I-PF-5-01].
\end{acknowledgements}

%% REFERENCES

%% The reference list is compiled as follows:

\bibliographystyle{copernicus} 
\bibliography{references}

%% GLOSSARY

\printglossary[type=main, title=Glossary]

\end{document}